\documentclass[a4paper,10pt]{amsart}
\usepackage{fourier}
\usepackage{biblatex}

\usepackage{graphicx}%

\usepackage{amsmath,amssymb,amsfonts}%
\usepackage[foot]{amsaddr}

\newlength{\jfigwidth}
\newcommand{\jpgfig}[3]{\jdofig{#1}{#2}{#3}{Figures}{.jpg}}
\newcommand{\pdffig}[3]{\jdofig{#1}{#2}{#3}{Figures}{.pdf}}
\newcommand{\jdofig}[5]{
	\begin{figure}\centering\includegraphics[width=#3\jfigwidth]{#4/#1#5} \caption{#2}\label{fig:#1}\end{figure}
}

\begin{document}

\title[Disk-stacking models]{Disk-stacking models are consistent with Fibonacci and non-Fibonacci structure in sunflowers} 
\author{Jonathan Swinton}
\address{ORCID 0000-0002-3196-9653}
\email{jonathan@swintons.net}

\begin{abstract}
This paper investigates a model of plant organ placement motivated by the appearance of large Fibonacci numbers in phyllotaxis, and provides the first large-scale empirical validation of this model. Specifically it evaluates the ability of Schwendener disk-stacking models to generate parastichy patterns seen in a large dataset of sunflower seedheads. We find that features of this data that the models can account for include a predominance of Fibonacci counts, usually in a pair of left and right counts on a single seedhead, a smaller but detectable frequency of Lucas and double Fibonacci numbers, a comparable frequency of Fibonacci numbers plus or minus one, and occurrences of pairs of roughly equal but non-Fibonacci counts in a `columnar' structure. A further observation in the dataset was an occasional lack of rotational symmetry in the parastichy spirals, and this paper demonstrates those in the model for the first time.

 Schwendener disk-stacking models allow Fibonacci structure by ensuring that a parameter of the model corresponding to the speed of plant growth is kept small enough. While many other models can exhibit Fibonacci structure, usually by specifying a rotation parameter to an extremely high precision, no other model has accounted for further, non-Fibonacci, features in the observed data.  The Schwendener model produces these naturally in the region of parameter space just beyond where the Fibonacci structure breaks down, without any further parameter fitting. We also introduce stochasticity into the model and show that it while it can be responsible for the appearance of columnar structure, the disordered dynamics of the deterministic system near the critical region can also generate this structure. 

\end{abstract}
\maketitle 

%\listoffixmes
 \section{Introduction}
The visible spirals often formed by the pattern of seeds in the seedhead of the sunflower \textit{Helianthus annuus} are known as parastichies (Figure~\ref{fig:319MOSIpaper}).
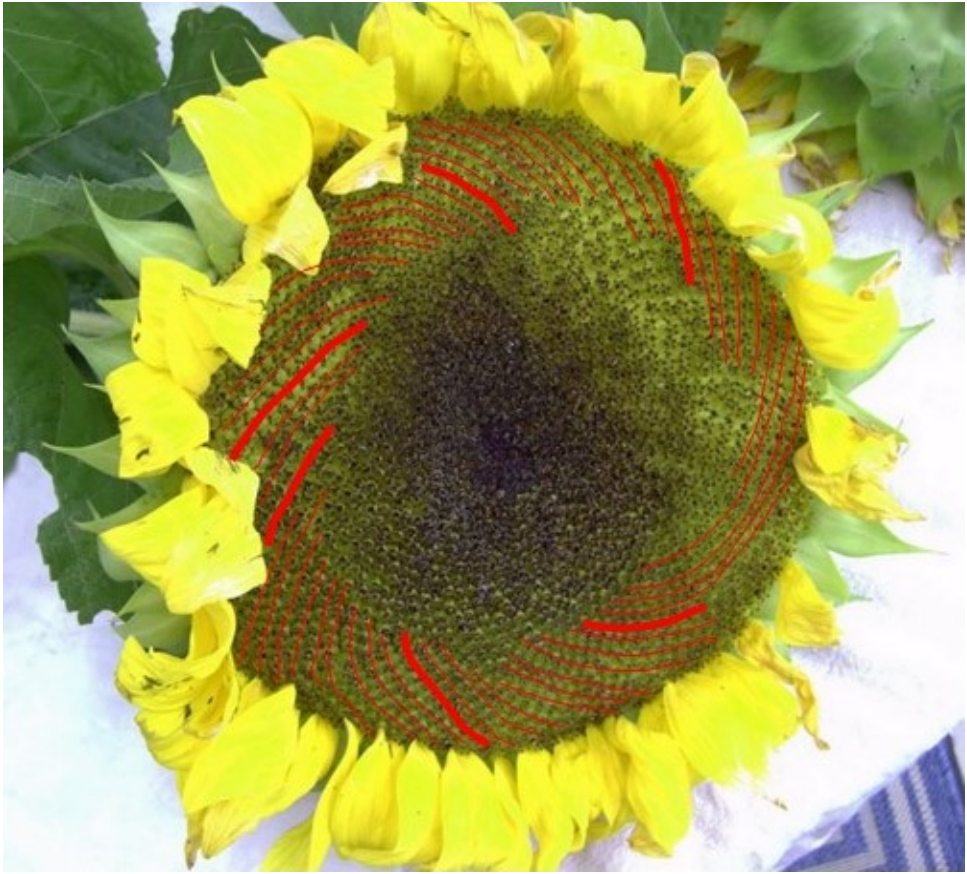
\jpgfig{319MOSIpaper}{A sunflower with a family of 53 parastichy spirals highlighted in red by a human annotator in one direction, but no clear family of spirals in the other. This photograph was taken after the small yellow true flowers that had earlier covered the seedhead, or capitulum, had set seed and fallen away.  Green, roughly triangular,  bracts remain half-visible around and below the outer rim of the capitulum, developmentally preceding the yellow ray-florets which in turn surround the individual seeds packed onto the capitulum base. In many large mature specimens, as here, towards the centre of the capitulum the seed-setting process has not completed and will not generate viable seeds. 
	Submission 319 of the MOSI dataset~\cite{swintonNovelFibonacciNonFibonacci2016}.}{1}
It has been observed for several centuries that the count of 53 parastichies in Figure~\ref{fig:319MOSIpaper} is relatively unusual: it is much more common to see counts like the Fibonacci numbers 34, 55, or 89 in a seedhead of this size. 

One cogent mathematical explanation for the appearance of this Fibonacci structure in the arrangements of plant organs more generally has been available since the 1990s~\cite{jeanSymmetryPlants1998}, and on some views since the 1900s~\cite{vanitersonjrMathematischeUndMikroscopischAnatomische1907} by considering the structure of the van Iterson tree, an organisation of lattice patterns on cylinders~\cite{godinPhyllotaxisGeometricCanalization2020,swintonMathematicalPhyllotaxis2023}. 

However the very assumption that plant organs \textit{are} placed in lattices has turned out to be too strong for the purposes of interpreting some of the data, and too abstract to be interpreted in the light of known biology which contains no global `protractor' mechanism to construct the lattices. 
By contrast, the modern molecular biology of organ placement gives a central role to auxin concentration as an initiator, with spatial structure evolving through auxin synthesis within recent new organ centres, followed by diffusion or more importantly active transport across cell walls, particularly by the PIN-1 family of proteins. Control of these initiation and transport processes has been modelled from a variety of influences including physical strain, auxin concentration itself, and genetic signalling and the individual cell level~\cite{godinPhyllotaxisGeometricCanalization2020}. Broadly speaking, modern molecular biology supports Snows' rule: new organs are created locally, one by one, as soon as there is space for them. 

In response to the limitation of lattice models, disk-stacking models have been recently revived~\cite{atelaDynamicalSystemPlant2002}. Although they date from the nineteenth century they have only recently become a significant paradigm for pattern formation~\cite{godinPhyllotaxisGeometricCanalization2020,goleFibonacciQuasisymmetricPhyllotaxis2016}. Because disk-stacking models allow lattice solutions,  the van Iterson paradigm remains a powerful organising principle for their dynamics, and it is by now well established that disk-stacking models can indeed demonstrate Fibonacci structure~\cite{goleFibonacciQuasisymmetricPhyllotaxis2016}. Moreover, as the van Iterson paradigm suggests, there are parameter regions in which strict Fibonacci patterns are lost but patterns remain closely ordered with either Lucas numbers or double-Fibonacci pair counts occurring~\cite{goleFibonacciQuasisymmetricPhyllotaxis2016,yonekuraMathematicalModelStudies2019}. An interesting possibility, not obvious within the van Iterson paradigm and only recently reported in the mathematical literature, is of paired but approximately equal parastichy counts~\cite{goleFibonacciQuasisymmetricPhyllotaxis2016}; these arise from a loss of the strong left-right asymmetry implied by Fibonacci structure. 

The main purpose of this paper is to confront disk-stacking models with the largest published dataset on parastichy counts in sunflowers~\cite{swintonNovelFibonacciNonFibonacci2016}; this allows us to  concentrate not on \textit{whether} Fibonacci structure can occur, but on understanding the parameter regimes which allow this and whether departures from Fibonacci structure can be caused by model dynamics or other noise. 
As a subsidiary aim, we also explore how a breakdown can occur of the spiral structure that allows biologically relevant parastichy counts to be made at all, and compare this with some observations from the same dataset.

\section{The van Iterson bifurcation tree}
  Simple models for the placement of seeds and other organs on the plant stem can be characterised either by a characteristic length scale, or in older models a combination of a characteristic rotation angle (the \textit{divergence}) and a displacement up the stem (the \textit{rise}) as in  Figure~\ref{fig:scpNonOpposed}). In either case, a mathematical consequence is the possibility of regular lattice patterns, within which straight lines called \textit{parastichies} can be defined which are a good model for the visually apparent spirals of botanical form.
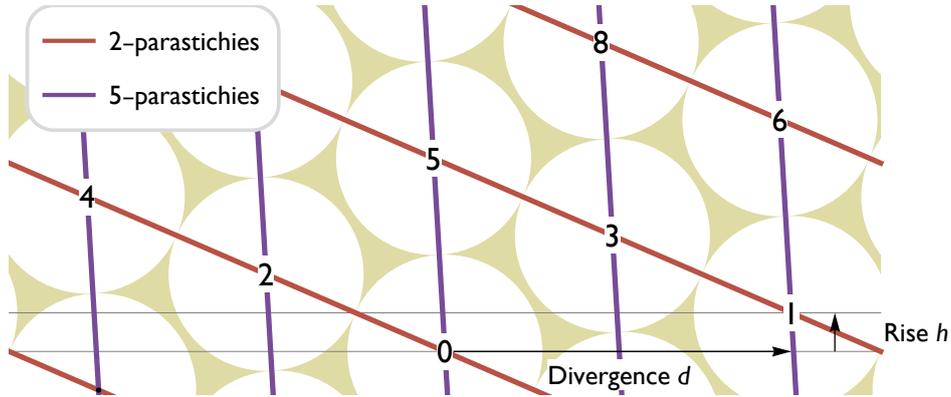
 \pdffig{scpNonOpposed}{A disk lattice, generated by a divergence $d$ and rise $h$ between each point and the next higher point on a vertical, horizontally-periodic cylinder. Each of the sides of the lattice tile are the same length; these sides form two families of parastichy lines which each spiral around the cylinder. there are two red lines and five blue lines intersecting any horizontal line, corresponding to a parastichy count of $(2,5)$. Equivalently, the shortest vectors in the lattice from point 0 are to points 2 and 5, where the points are numbered in increasing height order. Here, unusually, both parastichy lines point upwards and to the left: they wind in the same, rather than in  opposite, directions around the cylinder.  These \textit{non-opposed} lattices play an important role in understanding the structure of van Iterson space, but cannot be generated by disk-stacking models.
 }{1}%

  There is an elegant mathematical  classification of all such lattices on the cylinder through what is known as the van Iterson diagram~\cite{vanitersonjrMathematischeUndMikroscopischAnatomische1907,swintonMathematicalPhyllotaxis2023}, illustrated in Figure~\ref{fig:scpVanIterson}.  The  van Iterson diagram is a tree within this space of lattices, on which the lattice is a \textit{disk lattice}, in which the two shortest vectors in the lattice have the same length and so a disk with that length as the diameter can be placed at each lattice point.  Each branch of these possible disk lattices on the cylinder can be labelled by a pair of spiral counts, expressed as a pair of integers, the \textit{parastichy numbers}. The tree is formed by multiple bifurcation points corresponding to the hexagonal lattices at which it is possible for the parastichy numbers to change. Moving downwards through this tree the nature of the bifurcation is always the same and of the two choices,
that which corresponds to a closer-packing lattice will preserve Fibonacci structure. The central mathematical idea is to find a model of lattice formation in which what Turing in the 1950s called the `hypothesis of geometrical phyllotaxis' holds true: that the model always has this Fibonacci choice as the stable branch at the bifurcation~\cite{swintonTuringMorphogenesisFibonacci2013}. 
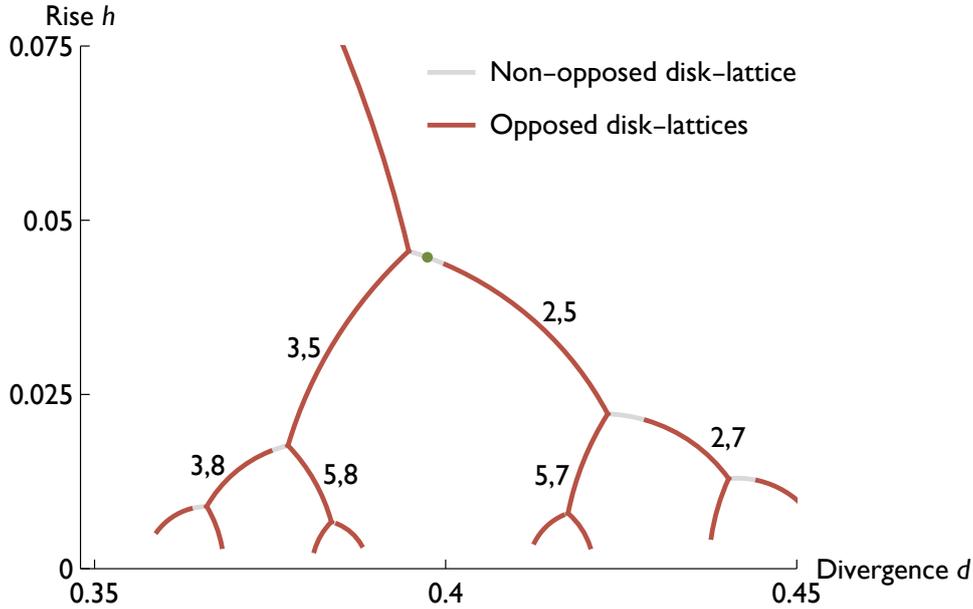
\pdffig{scpVanIterson}{The van Iterson bifurcation diagram for disk-lattices on cylinders showing transitions between different parastichy counts as a function of lattice divergence $d$ and rise $h$. Red lines show branches on which the lattice is an opposed disk-lattice, with parastichy lines going in opposite directions around the cylinder. Gray lines show branches with a non-opposed disk-lattice. Branches are labelled by their parastichy counts.  The green dot corresponds to the lattice parameters for the non-opposed lattice of Figure~\ref{fig:scpNonOpposed}.}{1}%

These mathematically defined parastichy counts, which label lattices by pairs of integers, are an attractive model for the empirical enumeration of spirals visible to the eye on the stem of many plants and, as we focus on in this paper, on that portion of the stem which is flattened into a disk-like cap to form the seedhead, or capitulum, of the sunflower. 
    Moreover, the ratio of the organ-formation length scale and the circumference of the stem  during development is a tempting candidate for a bifurcation parameter, in which maturation of the plant corresponds to moving through the van Iterson tree along increasing Fibonacci pairs. 
 By itself, though, the van Iterson tree cannot explain what is meant by a bifurcation between lattice-like patterns, and by construction it cannot exhibit non-lattice patterns. Both of these difficulties can be overcome by adopting disk-stacking models, which have lattices as a particular though not globally stable solution.

\section{Disk-stacking models}

 Disk-stacking models were first introduced by Schwendener in the 1870s as an exploration of organ placement in plants~\cite{schwendenerMechanischeTheorieBlattstellungen1878}.  Schwendener sketched the patterns seen when disks of decreasing size were stacked, one after the other, around a cylinder (Figure~\ref{fig:schwendener1878}).%
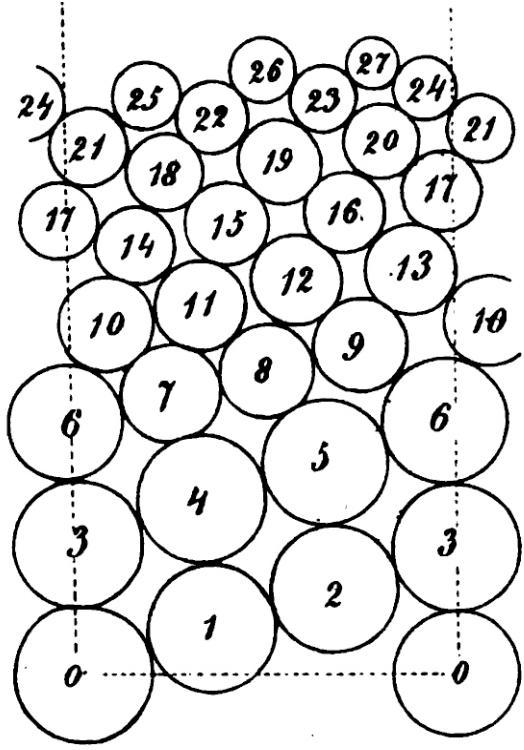
\jpgfig{schwendener1878}{Figure 31 of Schwendener, 1878~\cite{schwendenerMechanischeTheorieBlattstellungen1878}. 
}{.6}
Although Schwendener's work was not influential on mainstream plant morphology, it did influence the better known 1907 PhD thesis of van Iterson who explored the possible patterns arising from arrangements of fixed-size disks as a function of disk size and first drew a version of the van Iterson diagram~\cite{vanitersonjrMathematischeUndMikroscopischAnatomische1907,	schouteUberPseudokonchoiden1913}.
 The 21st century has seen a modest revival of interest in the dynamics of these stacked-disk models~\cite{
	atelaRhombicTilingsPrimordia2017,
	hottonPossibleActualPhyllotaxis2006,
	atelaGeometricDynamicEssence2011};  this paper was particularly strongly influenced by~\cite{goleFibonacciQuasisymmetricPhyllotaxis2016} and~\cite{goleConvergenceDiskStacking2020}. 	

Schwendener-type models are conceptually simple and fairly straightforward to implement (Figure~\ref{fig:scpDeterministicTransition}).
 We take a vertical cylinder with circumference fixed at 1. Apart from the initial conditions, the deterministic model is entirely defined by the function $r_i$ which gives the radius of the $i$th disk.
 For the models in this paper, we take  $r_i=r(\max_{j<i} z_j)$ to be  a function only of the height of the centre of the highest (and usually the most recently placed) existing disk
 For the tabulated results on parastichy counts, we take  $r(z)$ to be a piece-wise linear function of $z$ with $r(z)=r_L$ for $z<z_L$, $r(z)=r_U$ for $z>z_U$ and $r$ linearly interpolated between $r_L$ and  $r_U$ in the transition region from `Lower' to `Upper'. We use the slope of this linear interpolation, $r'= -(r_U-r_L)/(z_U-z_L)$, corresponding to the speed at which the disk size changes and chosen to be positive when the disk size is decreasing, as a control parameter in the simulations below.
\pdffig{scpDeterministicTransition}{
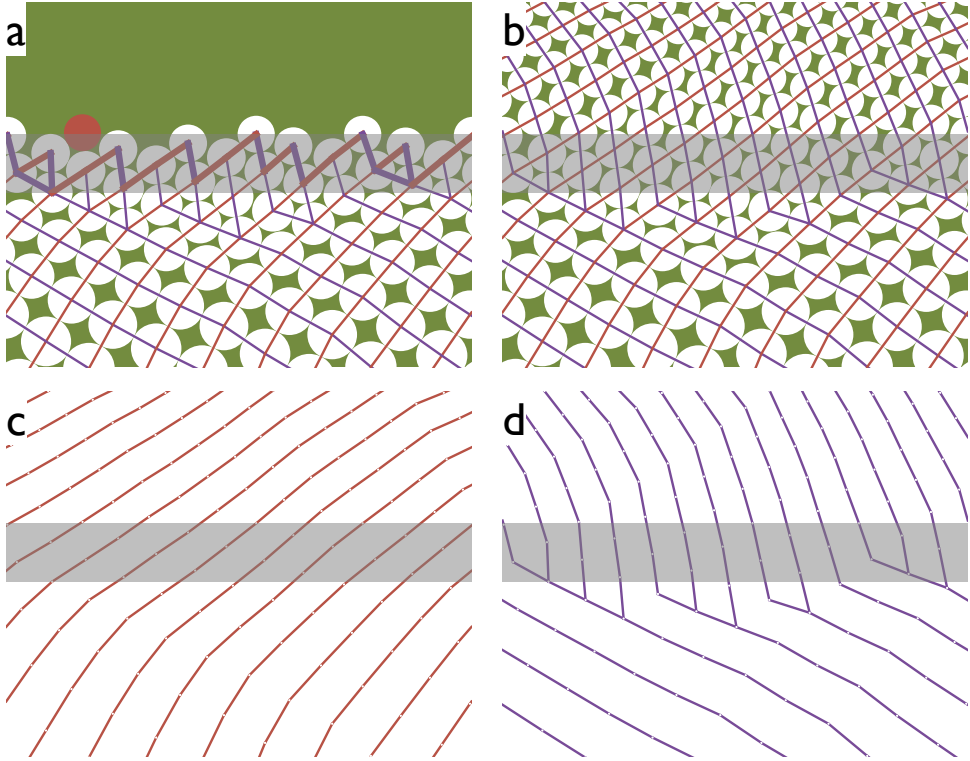
	Stacked disk models with varying disk radii enable transitions in parastichy counts. (a): Intermediate stage of a disk stacking model started near a square  (5,8) lattice. Disks are joined by lines to those they are in direct contact with, and the lines coloured red or blue depending on whether they slope upwards to the right or left. The position of the next disk to be placed (solid red) is determined only by the disks within the grey rectangle, and with the corresponding chain of connections shown with a thicker line. (b): completed model run, showing parastichy lines joining disk centres and a transition  from a (5,8) to an (8,13) lattice-like pattern. (c) right parastichy lines only, showing that a parastichy count of 8 is conserved throughout the run.  (d)  left parastichy lines only, showing how a series of $\gamma$ dislocations correspond to increases in the parastichy count from 8 up to 13. These transitions correspond to non-linearity of the topmost chain, as highlighted in (a). Diagram (a) also illustrates the Douady-Gol\'e method of parastichy assignment:  when the chain in (a) is traversed around the cylinder there are always 8 distinct sets of blue segments in the path, corresponding to the unchanging 8-parastichy of the red spirals in (c). By contrast the number of distinct sets of red segments can vary depending on the traversal taken, which allows this count to increase from 5 to 13 over the course of the run. In the numerical results the parastichy count for each disk is derived from the highest traversal through but not above that disk.
}{1}%  
 
  In order to introduce noise with magnitude $\sigma$ we first compute the position of the $i$-th disk using the deterministic model, but then multiply its radius by a scaling drawn from a random distribution which is uniform on $[1-\sigma,1+\sigma]$. It is this randomly rescaled disk which is then used to compute the position of subsequent disks. Disks are treated as `touching' for the purposes of computing parastichies if the un-rescaled disks touch.

\subsection{Cylinder to capitulum mappings}

The output from disk-stacking models can be mapped from cylinders to any surface of revolution, as in Figure~\ref{fig:scpConeTransformation} which shows how  results of disk-stacking models can be compared with empirical data on spirals at the outer sunflower seedhead rim. 
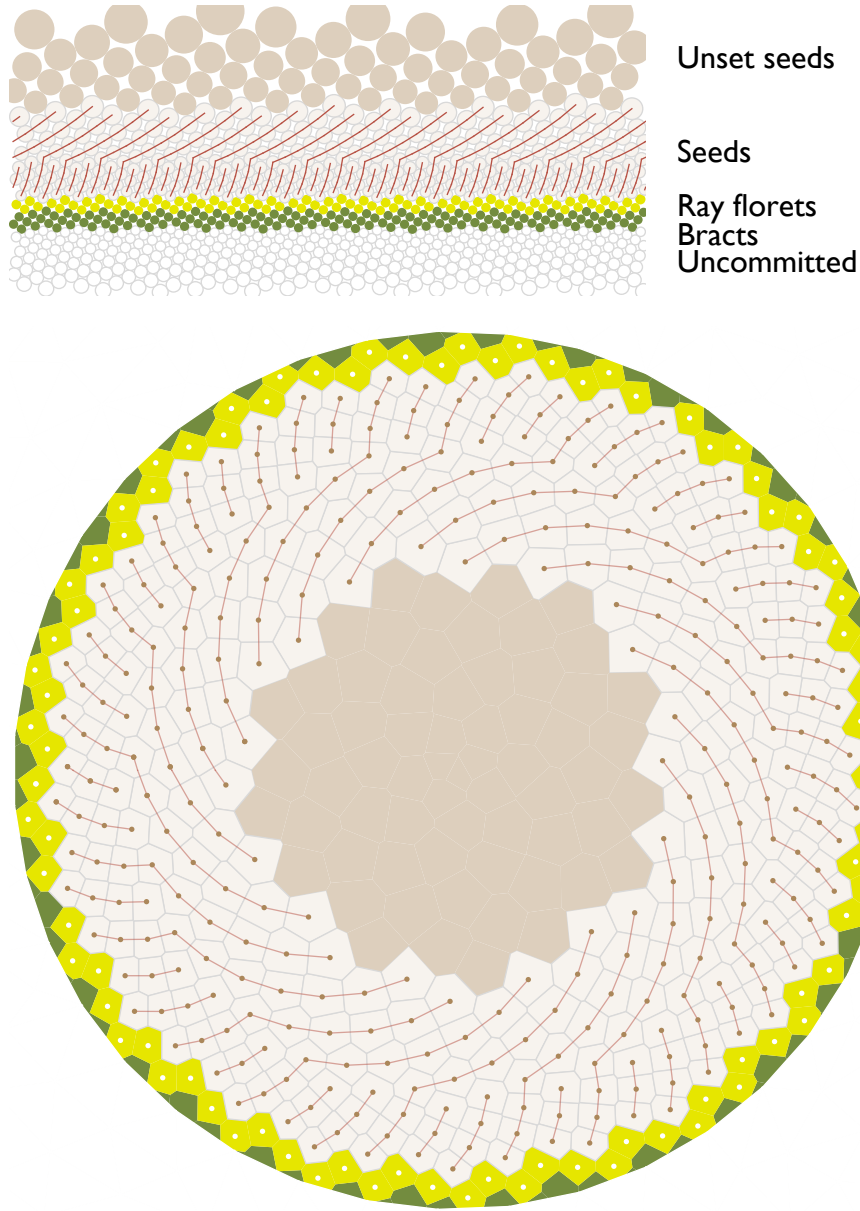
\pdffig{scpConeTransformation}{Mapping a disk-stacked pattern to a seedhead pattern. Top: a rising and then falling phyllotaxis modelled using a radius function whose inverse first linearly increases then linearly decreases; functional types corresponding to the colours and labels are arbitrarily added later to aid visualisation. After nodes labelled as uncommitted, bracts, and ray florets, and above an arbitrarily chosen point $z_S$, nodes are deemed to correspond to seed positions in the mature seed head.  Still higher node positions are deemed to correspond to a disk floret which does not proceed to set seed, as is common in the centre of large seedheads, up to the end of the run at $z=z_U$.   Right parastichy lines are shown in the region which will correspond to mature seeds in the adult seedhead. Below: the resulting placement pattern and parastichy lines mapped onto a seedhead disk. Points at coordinates $(x,z)$ on the cylinder are mapped to radial coordinates $\rho=(z_U-z)/(z_U-z_S)$, $\theta=2\pi x$. Note that occasionally some disks which are in contact in the cylinder pattern, for example in the ray florets), correspond to Voronoi cells which are not in  contact, partly because of the non-isometry of the mapping from cylinder to seedhead disk.}{1}

Crucially for the purposes of comparison to the empirical data, the principal parastichy counts are conserved by this mapping as long as it is not too distorting, which means that we do not have to address here the difficult biomathematical question of inferring the mapping from observational data. We do though note the dimensional argument that the disk-radius function $r(z)$ on the cylinder should scale like  the inverse of the circumference of the stem tip at the developmental stage when organ commitment takes place, together with the very crude assumption that this is in turn roughly proportional to $R(z)$, the mature stem radius, to find that roughly $r(z)\approx c/R(z)$. Observation on mature sunflowers suggests  a change in $R$ and hence $r$ of between 10 and 100 between stem and capitulum rim. 
We can compare this the results of Figure~\ref{fig:scpParastichyCountsTo89} below to see that, depending on the size of the initial disk radius, a scaling of  around 85 to 90 is needed to transition to a parastichy pair of (55,89). We might note that $89=F_{11}$, and more generally that van Iterson theory~\cite{swintonMathematicalPhyllotaxis2023} provides convincing arguments that if a transition to a Fibonacci pair $(F_n,F_{n+1})$ occurs, it is at a scaling of the disk radius of about $F_{n+1}$ from that of a $(0,1)$ lattice. 

We concentrate here on functional forms for $r(z)$ which decrease linearly to the point where the rim of the sunflower develops, corresponding to the data analysis of~\cite{swintonNovelFibonacciNonFibonacci2016} which counted the parastichy numbers at the outside rim of the seedhead. The majority of our numerical results are taken at the end of this linear  period, as indicated in Figure~\ref{fig:scpConeTransformation} by the points suggestively coloured for explanatory purposes as though they were bracts, ray-floret and seed positions, although our results do not rely on this classification.

\subsection{Parastichy numbers in stacked-disk models change at $\gamma$ and $\lambda$ dislocations}
 Parastichy numbers for the resulting patterns are assigned as follows, using Douady and Gol\'e's method~\cite{goleFibonacciQuasisymmetricPhyllotaxis2016}, and as shown in Figure~\ref{fig:scpDeterministicTransition}. This method coincides well with human assessment of spiral counts in relatively well ordered patterns but can still be assigned in strongly disordered patterns when the human eye is unable to identify structure. 
The mathematical definition of parastichy numbers to model spiral counts is well established for lattices~\cite{swintonMathematicalPhyllotaxis2023}, and gives a pair of integers which is fixed across the whole lattice. Stacked-disk models usually generate patterns which are not parts of lattices, but we look instead for a chain of touching disks that encircle the cylinder and return to the original.   A chain has a number of up-steps to higher disks and down-steps to lower disks and the pair of the total number of up-steps and down-steps is what we define as the parastichy count pair. The `top chain' for a disk is the unique chain that consists only of disks no higher than the starting disk and always choosing the higher of the possible next contact disks to the right. (In disordered regions there can occasionally be no contact disks to the right in which case the lowest of those to the left is chosen.)  Generally, we take the parastichy pair for a disk to be that of its top chain. 

If the pattern is exactly a lattice, or near to one,
then other chains  through a given disk will have the same parastichy count pair, but near pattern transitions, as illustrated in Figure~\ref{fig:scpDeterministicTransition}, the presence of triangles in the pattern graph give rise to  multiple possible  counts. It is in this way that the discrete-valued parastichy counts can transition by jumps of 1 as the disk-radius changes slowly. 
Each of these jumps corresponds to a `$\gamma$' dislocation (taking its name from the shape of the letter~\cite{zagorska-marekPhyllotacticPatternsTransitions1985}) of the parastichy lines,  as visible in Figure~\ref{fig:scpDeterministicTransition}(d). 
Conversely, $\lambda$ dislocations correspond to a decrease in the count.
\section{Observational data}
The reference dataset, and in particular Figure~\ref{fig:scpMOSIHistogram},  is taken from the MOSI Turing's Sunflower's project's 2012 collection of data on several hundred sunflower heads~\cite{swintonNovelFibonacciNonFibonacci2016}.    This histogram of parastichy counts, and other data in the Supplementary Information of that paper can be summarised as follows.
\begin{enumerate}
	\item There was a strong but not complete preponderance of Fibonacci counts.
	\item	Excluding Fibonacci counts, the next commonest parastichy count was one less than a Fibonacci number (specifically, 33, 54, or 88).  A Fibonacci number less one (like 33) was statistically significantly more likely to occur than a Fibonacci number plus one (like 35).
	\item The next most common was a Lucas number (29, 47, or 76), and then a double-Fibonacci number (42, 68).
	\item It was common to see sunflower heads in which parastichy spirals could be clearly counted in one direction but not in another
	\item In a small number of relatively small sunflower heads, pairs of nearly matching but non-Fibonacci parastichy counts like (11,11)  were seen.
\end{enumerate}

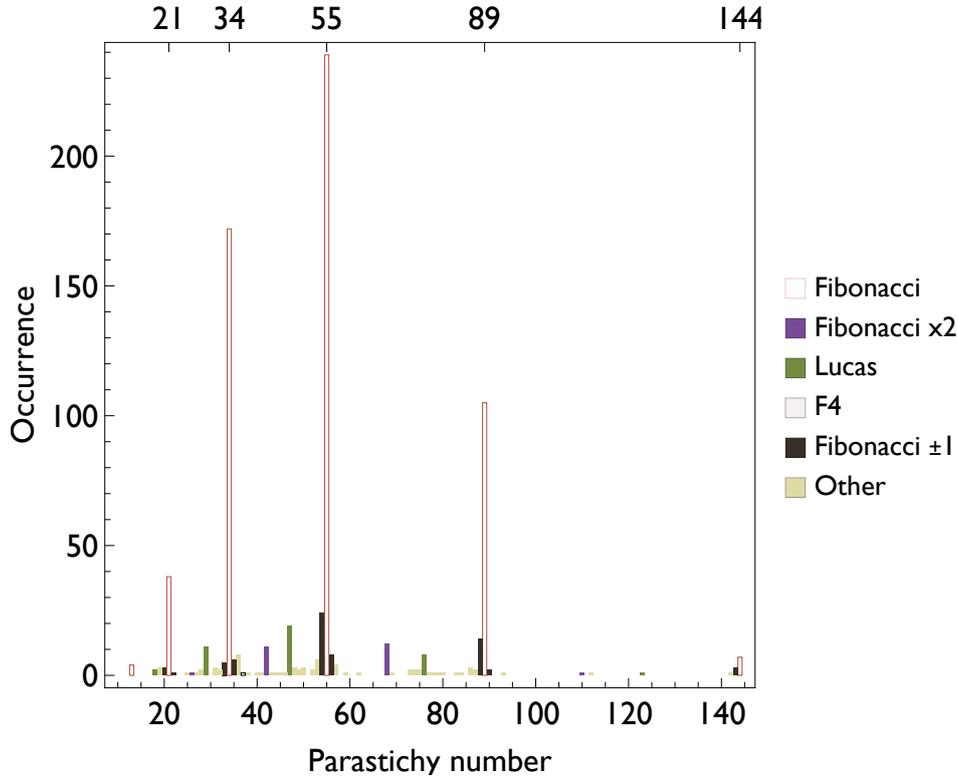
\pdffig{scpMOSIHistogram}{Counts of how often each parastichy number  was observed in the MOSI dataset. Redrawn from~\cite{swintonNovelFibonacciNonFibonacci2016}. An expanded view is shown in Figure~\ref{fig:scpAhaPair}.}{1}

\section{Results}

\subsection{Slow change rates support Fibonacci transitions}
	
Solutions to stacked disk models showing large Fibonacci pair parastichy counts were exhibited by Bursill and colleagues in the 1980s~\cite{jeanBib41,jeanBib721}.  Previous explanations for consistent choice of the Fibonacci branch in the van Iterson tree have either assumed that non-opposed lattices can never occur~\cite{mitchisonPhyllotaxisFibonacciSeries1977} or that they can be rejected as less-well-packed than the corresponding opposed lattice near the hexagonal bifurcation~\cite{douadyCh21Phyllotactic1998}. Because disk-stacking models cannot generically generate non-opposed lattices, they ought to display a preference for Fibonacci structure, at least as long as $r'$, the rate of change of disk radius with stem height,  is not too large. This was demonstrated in  numerical calculations of Gol\'e and colleagues~\cite{goleFibonacciQuasisymmetricPhyllotaxis2016}. We can see that Figure~\ref{fig:scpDeterministicTransition} reproduces these results by showing a (5,8) to (8,13) transition.
As $r$ continues to decrease, the process can be repeated, as illustrated in  Figure~\ref{fig:scpParastichyCountsTo89} which shows a series of transitions between adjacent Fibonacci pairs.  
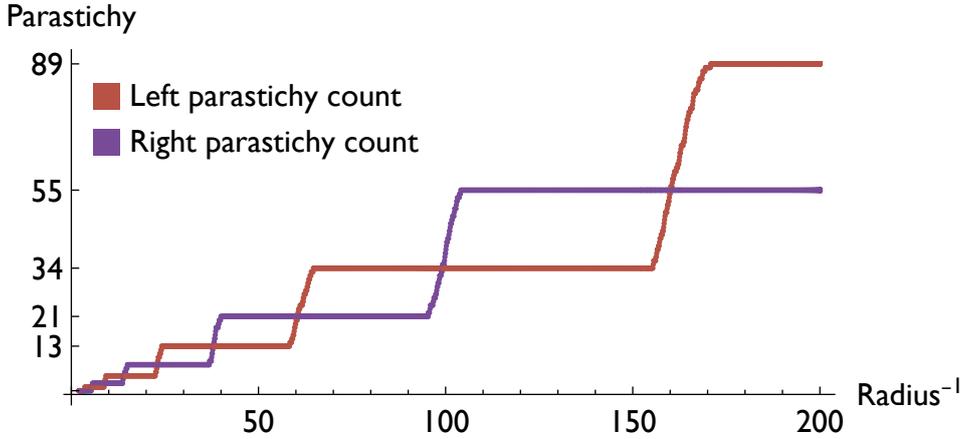
\pdffig{scpParastichyCountsTo89}{Transitions between parastichy counts as a function of the slowly changing inverse radius of the disks. $y$ axis: Local chain parastichy counts for the topmost chain through each successive disk in a single
	deterministic run with a  $r'=0.03$ started from a single disk of radius 1/2. $x$ axis: inverse radius of the disk at the highest point of that chain.}{1}

An understanding of the van Iterson tree adds further insight to these smooth, deterministic, transitions. In van Iterson parameter space, lattices change their parastichy numbers exactly when they are hexagonal lattices, then widen in internal angle into square lattices before narrowing again into hexagonal lattices at the next lower transition.  The tiles of Figure~\ref{fig:scpDeterministicTransition} reflect this transition. Because $r'$ is small, the disk pattern at the transition is very close to a hexagonal lattice, and the transition is able to occur smoothly. In particular, the positions of lattice points above and below the transition are strongly correlated: the change of parastichy number occurs by a smooth changeover between which of these points are closest to each other.

As expected from the van Iterson tree, this disk-stacking model can account for the dominance of Fibonacci counts in the empirical data of Figure~\ref{fig:scpMOSIHistogram}.  What it does not by itself account for is the other peaks in that data, which is the main concern of this paper.

\subsection{Parastichy count distributions after multiple transitions can match empirical data}
Figure~\ref{fig:scpAhaPair} illustrates the main result of this paper. It gives an example of the parastichy numbers observed with a single fixed $r'$ over the course of multiple stochastic replicates of a disk stacking model.    There are some qualitative differences at this particular example: the Fibonacci numbers are less overwhelmingly dominant than in the empirical dataset, while the Lucas numbers are more frequent, rather than less frequent, than the $F\pm1$ counts. Nevertheless these capture the key observations drawn from  Figure~\ref{fig:scpMOSIHistogram}: the dominance of Fibonacci numbers, the  presence of Lucas and double-Fibonacci numbers, and,  not previously demonstrated in models, the presence of Fibonacci numbers plus or minus 1. 
The parameters $r'=0.03$ and noise $\sigma=0.05$ used to generate Figure~\ref{fig:scpAhaPair} were not specified \textit{a priori}. They were chosen after inspection of the wider grid of Figure~\ref{fig:scpTo55} but there was no further tuning of the model;  indeed the model has no further free parameters to tune.

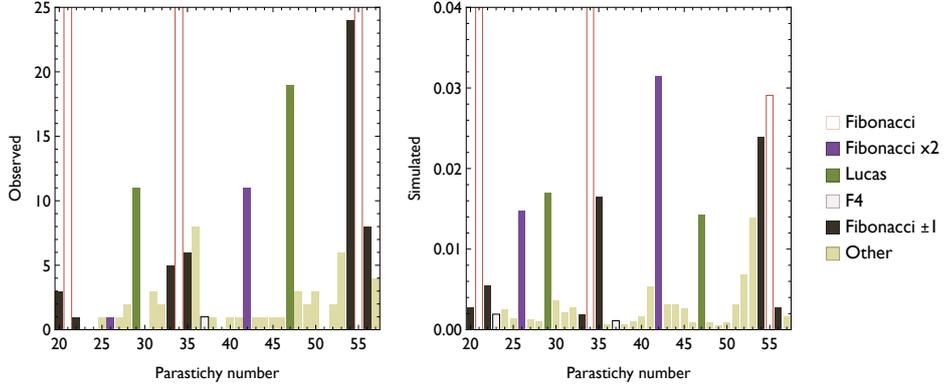
\pdffig{scpAhaPair}{Left, empirical parastichy counts redrawn from Figure~\ref{fig:scpMOSIHistogram} to emphasise non-Fibonacci observations. Right,  parastichy counts in a disk stacking model with $r'=0.03$ and noise $\sigma=0.05$. In each simulation the initial condition was a single disk of radius $r(0)=1/2$, with a disk size function $r(z)=1/2- r'z$, run until $r(z)$ changed by a factor of  60, corresponding in a lattice model to a change from a (0,1) lattice to a (34,55) one. 
	Simulated parastichy numbers were pooled over the course of each run and then further pooled over 10 replicates of the randomisation. }{1}

\subsection{Sensitivity of parastichy distributions to model parameters}

The two-dimensional parameter space arising from varying slope $r'$ and noise $\sigma$ was scanned as shown in  Figure~\ref{fig:scpTo55}. As expected, at slow disk change rates and low noise, a strong dominance of Fibonacci numbers 13, 21, 34 and 55 is observed in the upper left corner of the table. For rapidly changing and noisy simulations run over the same 60-fold reduction in disk radius, shown in the bottom right corner of the table, the 34 and 55 peaks completely disappear and there is a cluster of parastichy counts between 34 and 55, which appears to peak at the Lucas number 47, although as discussed below this is more a coincidence arising from the fact that 47 is close to the 44.5, the mean of 34 and 55, than a result of Fibonacci structure. 

Comparison of the first row (with low disk change speed $r'=0.01$) and the first column (with no noise: $\sigma=0$) suggests that the transition between these two outcomes can be achieved either by speeding up the disk radius change or by increasing noise, but these two mechanisms interact in a complex way: for example the $r=0.07$ row shows increasing noise first increases the prevalence of Fibonacci counts before then decreasing it. To understand this behaviour we need to look at the nature of the pattern transitions in more detail.

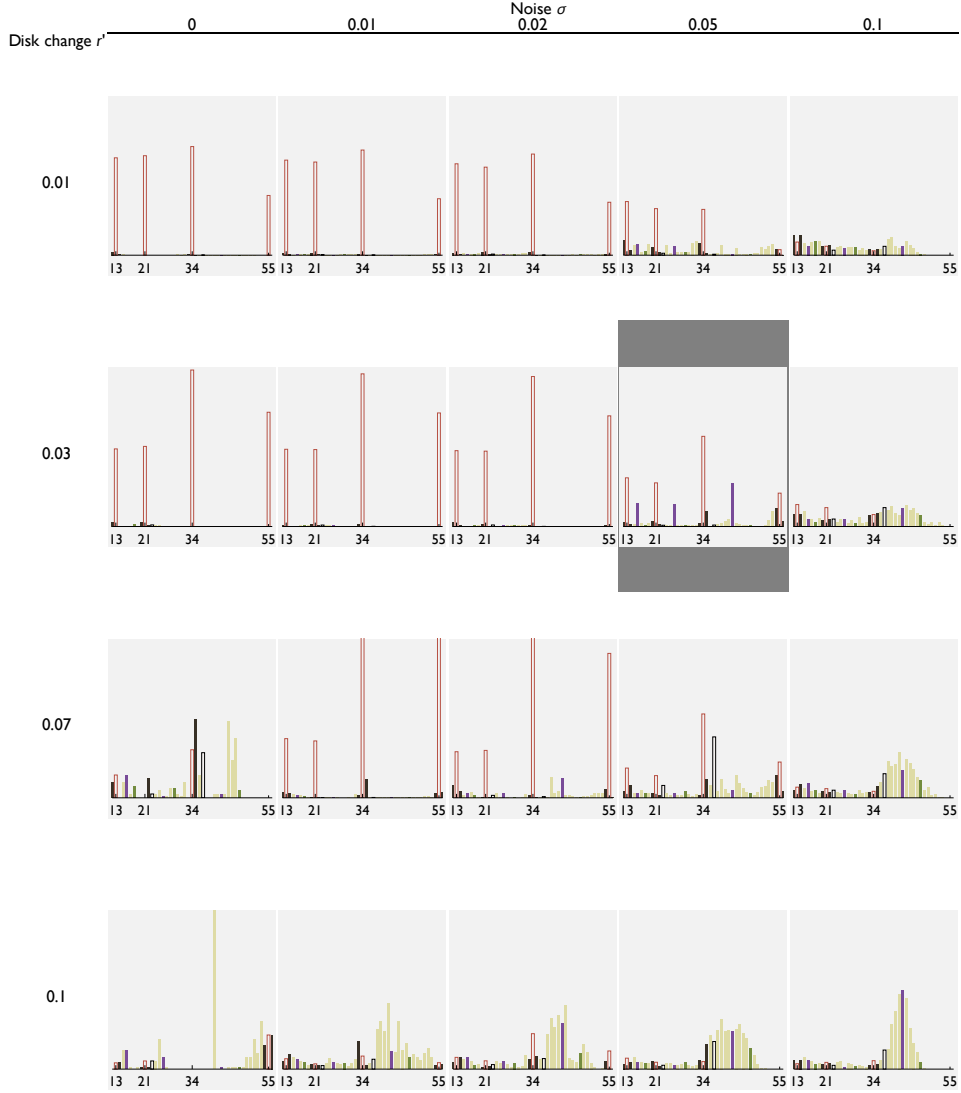
\pdffig{scpTo55}{Parastichy histograms depending on speed of disk size contraction $r'$ and magnitude of noise $\sigma$.  Runs and parastichy numbers colour-codings as in Figure~\ref{fig:scpAhaPair}. The histogram shown in Figure~\ref{fig:scpAhaPair} is highlighted. The vertical axis of each histogram is a relative frequency count; bars higher than 0.07 are truncated vertically.}{1}

\subsection{Deterministic transitions around criticality generate near-Fibonacci counts, but also introduce dynamic noise}

This section explores how a pattern starting from an exact $(8,13)$ lattice evolves: Figure~\ref{fig:scpDet813} illustrates some example runs.
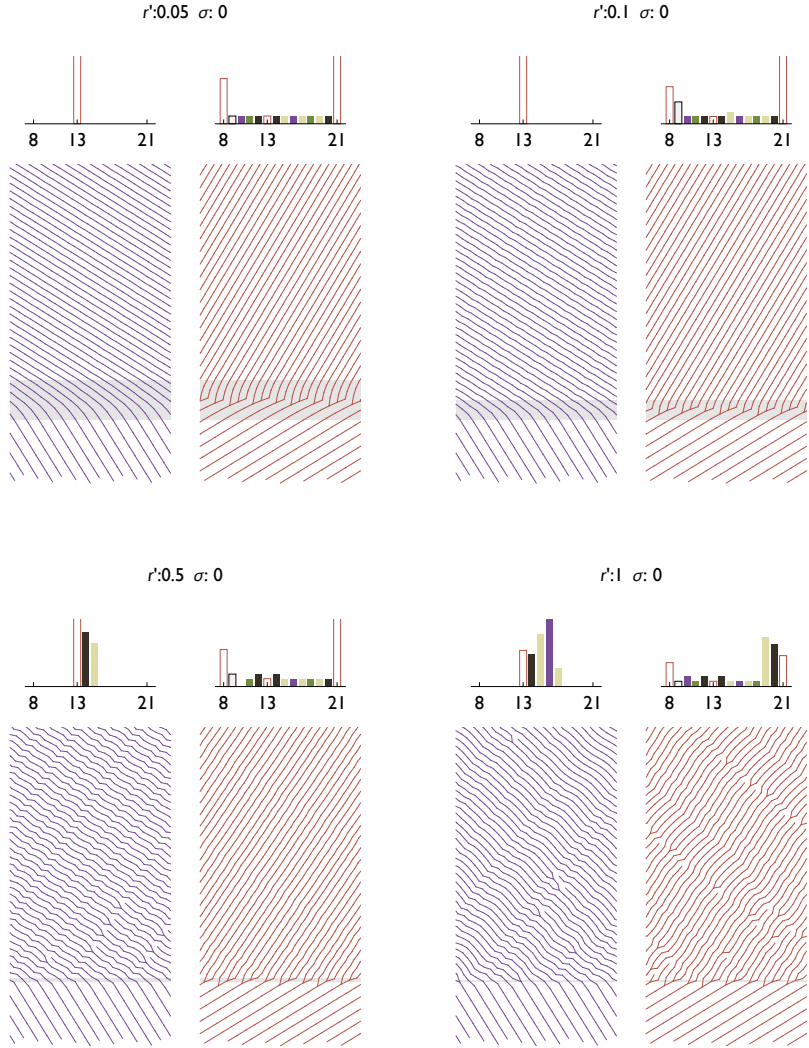
 \pdffig{scpDet813}{Deterministic dynamics starting from an (8,13) square lattice at different transition rates $r'=$0.05, 0.1, 0.5, 1.0 respectively. In each case the disk size is fixed outside of the transition zone marked with a gray rectangle, and linearly changed by a ratio of 1.6 within that transition zone as defined in the Method section with slope $r'$.  Histograms are of $\log_{10} (1+x) $ where $x$ is the binned parastichy count at each disk evaluated using the chain method of Douady et al, and coloured as in Figure~\ref{fig:scpAhaPair}. Below each histogram is the corresponding set of parastichy lines; changes in parastichy counts are associated with $\gamma$ and $\lambda$ dislocations in these patterns.}{.9}
 van Iterson lattice theory predicts that the gradient of what is initially the higher parastichy count lines, here the left-winding blue 13-parastichies, will gradually rotate through the transition zone. The gradient of the lower, 8-parastichy lines, which are the red, right-winding ones, also rotates slowly below the transition point, but then there is a sharp transition to the 13-parastichies. Moreover  van Iterson theory predicts that at that point the 8- and 13-lines are $120^{\circ}$ apart, and the new 21-lines bisect this angle and make an angle of $60^\circ$ with each of them, reflecting the hexagonal lattice underlying the bifurcation point. Before the transition the two angles each set of parastichy lines make with the vertical are asymmetric, with the blue lines steeper, while after they are also asymmetric but now it is the red lines which are steeper.
 
  In the transition region itself a series of dislocations in the red lines correspond to a rapid change of the red count from 8 up to 21, which can be detected in the 
 log histogram for the parastichy counts, but is always localised to the transition region. Turning to the 13-parastichies,  a further dynamic complexity emerges.
 For the  $r'=0.5$ panel of  Figure~\ref{fig:scpDet813} there is an occasional occurrence of 14 in the left winding count. Although the relative frequency of this $F+1$ count is low, it does intermittently occur, and occurs through occasional $\gamma$ dislocations away from  transition zone; by contrast the 21-parastichies remain conserved once past the transition region. 
 
At the yet higher value of $r'=1$, two further effects are visible in the left-winding parastichy lines. First of all,  the typical left parastichy count increases from 13 to 16; in addition dislocations in both the left- and right- winding parastichy lines continue over the  course of the entire simulation and produce a blurring of the peaks so that both 15 and 17 are seen in the left counts and 19, 20, and 21 in the right counts. All of these more complex change in parastichy counts occur outside of the transition zone and during the fixed-disk-size regime of the simulation.  In summary a form of deterministic noise is introduced into the dynamics of the model by a the jump to new and non-lattice conditions  in the narrow transition zone; near to the critical $r'$ value below which Fibonacci transitions occur this can promote $F+1$ counts.

\subsection{Rapid transitions promote near-Fibonacci counts while noise promotes columnar structures}
This numerical observation of deterministic chaos arising only at rapid transitions offers one intriguing potential explanation for non-Fibonacci structure. In order to explore whether this has a detectably different signature to the presence of biological noise we further explore the effect of non-zero noise $\sigma$ in the model.   
Figure~\ref{fig:scpSch813} is an analogue of Figure~\ref{fig:scpDet813},
 with the indicated level of noise.
We can see a number of the same phenomena as in the deterministic runs: a narrow transition zone still acts to perturb the initial condition of the post transition fixed-disk-size dynamics, with the corresponding shift of the parastichy counts. So for example, the $r'=1$, $\sigma=0.02$ case shows a predominance of the 16 and 19 counts analogously to the deterministic model. The same underlying deterministic dynamics also creates paired $\gamma$ and $\lambda$ dislocations that generate small amounts of blurring of the peaks.
 
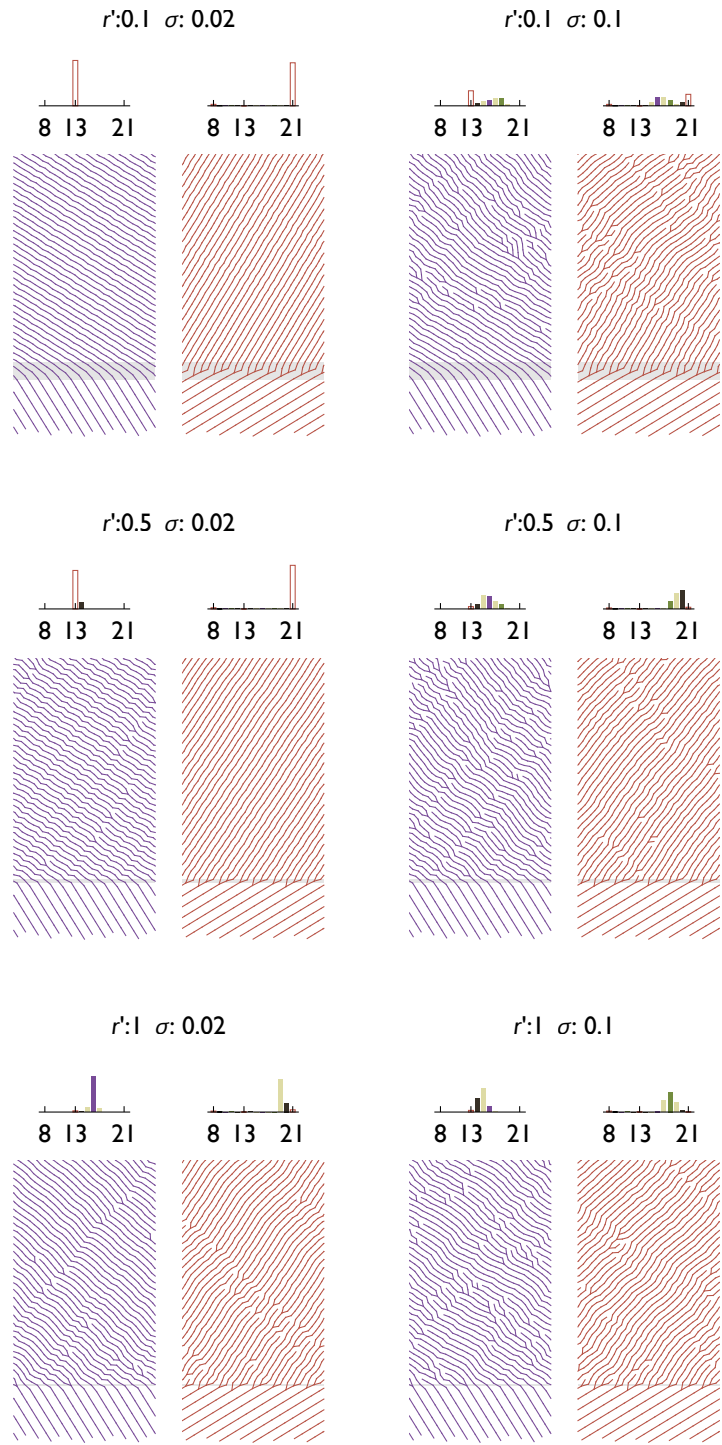
\pdffig{scpSch813}{Stochasticity flattens out parastichy count histograms, but also shifts them into the centre of the Fibonacci pair range. Single replicates of Figure~\ref{fig:scpDet813},  but with the addition of noise $\sigma=0.02$ or $\sigma=0.1$.} {.8}

The impact of noise is clearly seen in the increased number of dislocations in the parastichy lines. As specified in the methods section,  noise is implemented in such a way that disk positions are changed but disks are considered to remain in contact for drawing parastichy lines, so the contact lines are \textit{not} broken by this perturbation. Instead, additional dislocations can occur alongside those caused by the deterministic dynamics because the disk above the randomly-moved disk is in turn moved `out of position'. 

However there is an additional phenomenon which is not easily seen in the deterministic runs, but can be noticed in, for example the $r'=1$, $\sigma=0.1$ case. Looking at the gradient of the right-winding red parastichy lines after the transition zone, they share the gradient of the 21-parastichies seen in the less noisy case $\sigma=0.02$ and indeed correspond to parastichy counts of 21 in that region. However, starting some considerable time after the transition zone, there is a pattern shift and a series of unpaired $\lambda$ dislocations see a flattening of the gradient to about 45$^\circ$. The left-winding blue parastichies  also  become close to 45$^\circ$, although since they leave the transition zone already closer to 45$^\circ$ anyway the shift is less clear. 

What is happening in this new transition is a loss of asymmetry:  Fibonacci type structure requires that the one of the families of parastichy lines is much flatter than the other. In terms of disk placement this means that say, the disk above and to the left (say) of any given disk must be in general lower than the disk above and to the right of that disk. However with random initial conditions, or run stochasticity, there is no enforcement of this property, and there appears no theoretical reason why the dynamics of the disk-stacking model should tend to return the asymmetry. As a consequence, a generic run of the model from arbitrary initial conditions should be expected to yield parastichy lines with nearly equal angles, and correspondingly nearly equal parastichy counts. If, as in these simulations, the disk radius at the end of the simulation is calculated to allow a (13,21) pair, then this argument shows that two nearly equal parastichy counts adding to 34, in other words (17,18) are also likely patterns. Similarly, if the geometry is such that a (21,34) pair is possible then a generic pattern will be close to a (27,28) pair. These columnar solutions to the disk stacking model were identified by Gol\'e and colleagues~\cite{goleFibonacciQuasisymmetricPhyllotaxis2016} as QSS or `quasi-symmetric scenarios'  and an example is shown in Figure~\ref{fig:scpColumn}.
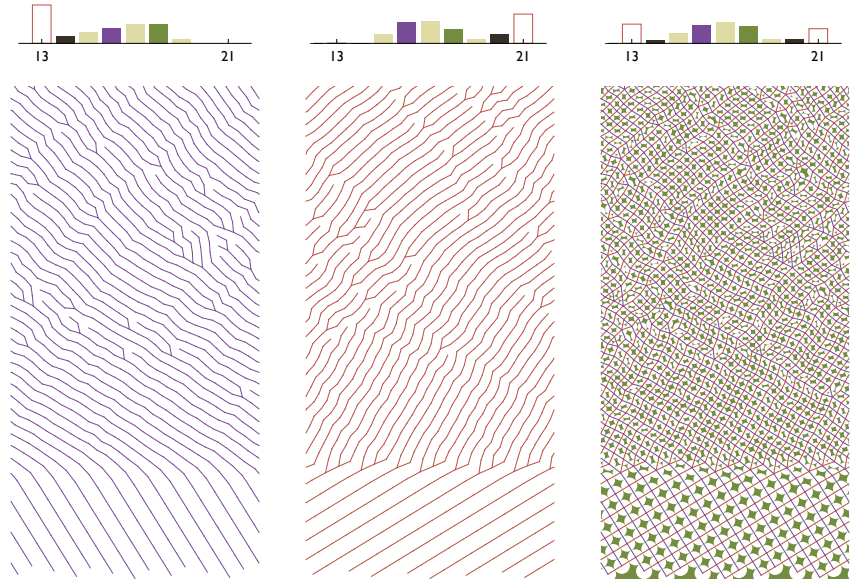
\pdffig{scpColumn}{Column-type patterns (and roughly equal parastichy pairs) emerge when noise is large enough to move away from Fibonacci structure. The resulting pattern contains patches in which the parastichy lines are each at $45^\circ$ and the disks are arranged in alternating vertical columns, which correspond to roughly equal parastichy numbers, here 16, 17 or 18. As Figure~\ref{fig:scpSch813} with $r'=0.1$ and $\sigma=0.1$, }{1}

\subsection{Double-Fibonacci and Lucas numbers emerge at large $r$ near the critical $r'$ and are promoted by noise}
The possibility of double-Fibonacci and Lucas numbers in disk-stacking models has already been demonstrated to occur near critical $r'$ values (e.g. Figure 15 of~\cite{goleFibonacciQuasisymmetricPhyllotaxis2016} or Figure 5 of ~\cite{yonekuraMathematicalModelStudies2019}.) However these results by themselves do not indicate at which point in pattern evolution these shifts from exact to more general Fibonacci structure occur. 

van Iterson theory shows a strong geometrical constraint on the parastichy numbers near lattice transitions. If the parastichy numbers before the bifurcation are $m$ and $n$ with $m<n$ then then the lattice geometry enforces that the higher parastichy number after bifurcation must be $m+n$; depending on the choice of branch in Figure~\ref{fig:scpVanIterson} the lower parastichy number is either $m$ or $n$, and only on the opposed branch is it $n$ so that the new pair is $n,n+m$. This branch choice is enough to enforce successive Fibonacci parastichy pairs from a starting pair of $(1,2)$, but it has long  been noticed that it also enforces successive Lucas pairs from a starting pair of $(1,3)$ and successive double-Fibonacci pairs from a starting pair of $(2,4)$, and that as we saw in the MOSI dataset such counts are frequent amongst non-Fibonacci observations. van Iterson lattice theory, though, cannot readily explain why and when these jumps which  ignore the normal branch choice occur.  If  it is possible to jump across the van Iterson tree once why not many times? In other words, suppose a Lucas pair $(11,18)$ is observed. Has that indeed evolved from an initial $(1,3)$ pattern? Or is it more likely for the jump to occur later so that, e.g., a $(8,13)$ pair directly transitions into an $(11,18)$. 
Disk stacking models support the hypothesis that only early transitions are likely: if either noise or too-fast transitions are strong enough effects to shift away from the left-right asymmetry of the  Fibonacci structure, the most likely resulting patterns are the columnar structures of the previous section, and these are close to Lucas or double-Fibonacci patterns only at the beginning of pattern formation.

\subsection{Disk stacking models generate falling phyllotaxis and observed asymmetry on the capitulum}

The observational approach adopted in the 2016 analysis of the MOSI images was to visually identify spirals as far out on the capitulum as possible, and so for the primary analysis, Figure~\ref{fig:scpMOSIHistogram} it has been enough to consider only rising phyllotaxis, corresponding to decreasing $f$. 
However when analysing the MOSI images, we did call attention to a previously unreported departure from Fibonacci structure, which we there called a lack of rotational symmetry~\cite{swintonNovelFibonacciNonFibonacci2016}. Perhaps surprisingly, the stacked-disk model for falling phyllotaxis, with increasing $f$ is also capable of illustrating this type of pattern, as illustrated in Figure~\ref{fig:scpFalling}, which 
shows such a simulation, together with a mapping of the simulation back to the capitulum surface as described in the Methods section. 
For rising phyllotaxis, the transition zones where parastichy number change are very narrow and horizontal, even in the highly-disordered example of Figure~\ref{fig:scpColumn}. By contrast, the transition zone in this falling phyllotaxis example is not horizontal, but follows the gradient of the parastichy lines and so  there is a large region of the cylinder on which the parastichy counts are difficult and perhaps meaningless to assign, Looking at the lower right panels we can see how the uncountability manifests itself: there is a family of 21-ish red contact lines making one angle in the cylinder, and another family of 8 red-contact lines making another, but the two families coexist over an extended region of the cylinder. This is the pattern that was identified in sample 667 of the MOSI image, whose hand-assigned parastichy lines are reproduced in Figure~\ref{fig:667nophoto}.
 
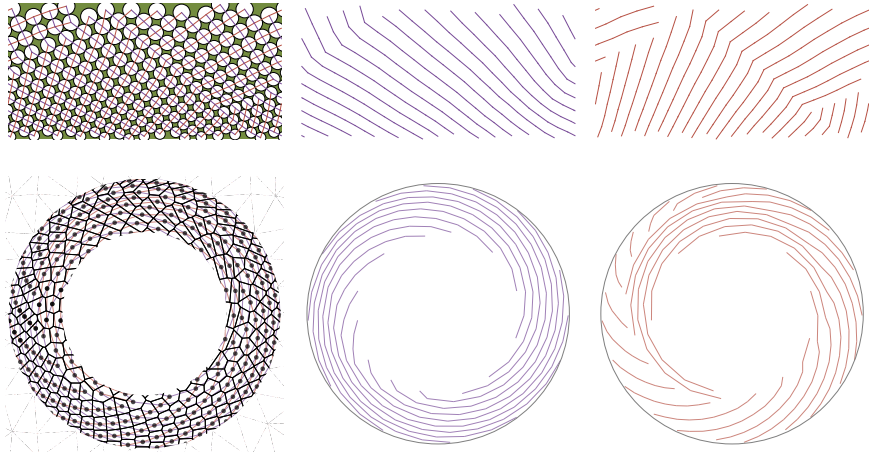
\pdffig{scpFalling}{Stacked-disk models on the cylinder can generate radially asymmetric patterns by rising phyllotaxis to the capitulum rim followed by falling phyllotaxis on the capitulum. This simulation shows horizontal variability in the location on the cylinder of dislocations of the red right-ward leaning parastichy lines. When the cylinder is mapped to a disk, this corresponds to a visible rotational asymmetry on the simulated capitulum.    There is no satisfying way of assigning `spiral counts' on the outer rim for the red contact lines: compare Figure~\ref{fig:667nophoto}.}{1}

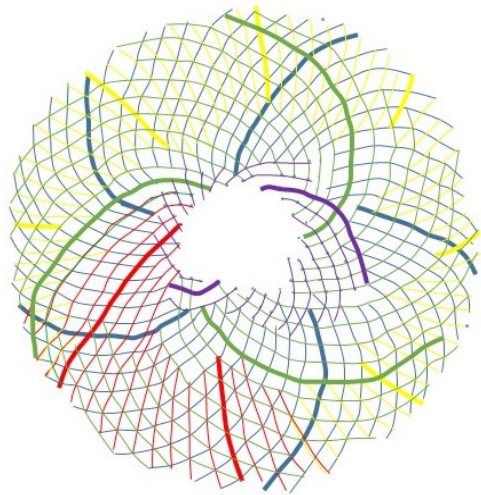
\jpgfig{667nophoto}{Observed contact line patterns in sample 667 of the MOSI dataset, with no satisfying way to assign a spiral count to the family of red/yellow/green spirals reflecting a rotational asymmetry as modelled in Figure~\ref{fig:scpFalling}.}{.5}

\section{Discussion}
It was the main motivation of this paper to take the implicit mathematical properties of stacked-disk models, notably those listed in~\cite{goleFibonacciQuasisymmetricPhyllotaxis2016} and make them explicit in a numerical example which could be compared with data, and to that end Figure~\ref{fig:scpAhaPair} is the main result. This Figure confirms that the dynamics of slowly changing stacked-disk models naturally generate well-ordered lattice-like patterns with Fibonacci structure, organised through the van Iterson tree, and provide a rationale for Turing's `hypothesis of geometrical phyllotaxis'~\cite{turingMorphogenTheoryPhyllotaxis2013,swintonMathematicalPhyllotaxis2023}. But it also shows that, for faster rates of disk change, different parastichy counts emerge in ways compatible with observed data.  Models built within this framework can generate plausible simulations of sunflower head patterns, as in Figure~\ref{fig:scpConeTransformation}, with  only one weak assumption, that the rate of geometric change is slow.  Even departures from rotational symmetry can be accounted for within the deterministic model behaviour rather than needing to invoke noise.
Taken together, this is a compelling mathematical story, and while a handful of other mathematical explanations for Fibonacci structure persist~\cite{okabeUnifiedRulePhyllotaxis2019}, it is the only one that has any support in the contemporary molecular biology of stem development~\cite{swintonMathematicalPhyllotaxis2023}. Mathematical explanations for Fibonacci structure based on `optimal packing' or a molecular protractor capable of controlling mean divergence angles to multiple decimal places should now finally be retired.

Intriguing mathematical questions abound about this rich class of models, although there are very few analytical results~\cite{goleConvergenceDiskStacking2020}. There is at present no theoretical guidance on the critical value $r'$ for the loss of Fibonacci structure, nor the interaction between this and the magnitude of the noise $\sigma$.  We have poor tools to understand  long-term behaviour, and it may well be that paradigms from, e.g., the statistical physics of dislocations will be helpful; in  particular it would be useful to understand the relaxation rate, if any, to asymptotic behaviour.  There is a deep connection between the structure of the van Iterson tree and the action of SL(2,Z) on the hyperbolic plane, and speculatively it is not impossible that this can be exploited for these more general models~\cite{swintonMathematicalPhyllotaxis2023}.  

This paper has purposely not tried to fit individual seed positions to any model.  Disk-stacking models have already been shown to be capable of this in, for example, individual pine cone examples~\cite{douadyFibonacciQuasisymmetricPhyllotaxis2016} or sunflower examples~\cite{hottonPossibleActualPhyllotaxis2006}. These studies have been remarkably fruitful in identifying new pattern structures both in models and observation and lend strong support to the paradigm, but are limited by data availability.  Advances in automated image analysis are likely to change this in the near future, and this will introduce a further mathematical and statistical challenge in fitting this high dimensional but strongly correlated data, not least because of the non-isometric mapping between cylinder and stem surface. Pattern identification tools that go beyond parastichy counts will be particularly useful.

Moving to outstanding biological questions, it is important 
to acknowledge that disk-stacking models make strong and easily refutable assumptions, such as 'consider a perfectly circular exclusion zone'. Together with the difficulties involved in mapping the actual developmental geometry of a cellular architecture to a mathematical cylinder it is not a realistic goal to `estimate $r=f(z)$' and exhibit simulations in which the model parameters are independently estimated from cellular kinetics. Nevertheless van Iterson theory does make predictions about how the scale of the apical meristem must change for each shift in parastichy pair and these predictions are amenable to testing. Furthermore this paper shows that the models can be tested well away from the idealised Fibonacci regime: unusual meristem dynamics, such as the bidirectional flowering of the teasel \textit{Dipsacus fullonum}, could  valuable independent demonstrations of the model's ability to link developmental scaling to mature morphology~\cite{naghilooUnderstandingUniqueFlowering2017}.

Another prediction arises from the  feature of mature sunflower morphology that the visible stem generally has no obvious spiral parastichies at all, although they can in some cases be identified~\cite{couderInitialTransitionsOrder1998}. The underside of the capitulum, where the largest changes in parastichy counts would be expected to be taking place, has no visible organs to be counted. A strong prediction of this paradigm is that the presence of large Fibonacci pairs in parastichy counts cannot occur de novo but must have been preceded by lower Fibonacci pairs, and should be possible to detect this pre-structure, either molecularly or through microstructure, in the underside of the mature capitulum or during its development. Another
potential prediction comes from panel a of Figure~\ref{fig:scpDeterministicTransition} where the highlighted chain can be thought of as defining the inhibition contour around the apical meristem. It is an essential feature of the paradigm that in any one case this contour is not left-right symmetrical: the red lines must be on average flatter than the blue ones; detecting this asymmetry in practice might be  hard, but it would be  a key signature for a molecule claimed to be essential in the organ generation and inhibition process. An illustration of the value of the stacked-disk paradigm is that it further predicts that if it is possible to detect asymmetry in some cases and not others, then the symmetric cases will be associated with rapid geometric change in general and non-Fibonacci counts in particular. 

The significance of this paradigm goes well beyond the sunflower capitulum, or any of the other notable examples of large Fibonacci pairs in plant form.  The presence of this Fibonacci structure is a powerful clue to the possible mechanisms but the mechanisms themselves are very unlikely to be specific to these cases. Node placement mechanisms came early in the evolution of the stem and the variety of outcomes possible from disk-stacking models may well be relevant across a wide range of plant architectures. Modern developmental biology has yielded a  deep, if still incomplete, understanding of the genetic and biochemical control of stem morphology, together with molecular probes for the intensity of important signalling  molecules such as auxin, and the ability to visualise the dynamics of the process over time in 4-dimensional confocal microscopy.  Yet the quarter century since the mathematical paradigm of the van Iterson tree was worked out has not seen any mutually productive confrontation of these paradigms with such rich data, nor attempts to exploit them in relation to more detailed mechanistic and cell-based models~\cite{newellFibonacciPatternsCommon2013}. This paper was written out of a conviction that there is `no gene for the number 55': the observed structural motifs of phyllotaxis can \textit{only} be understood in the light of a mathematical understanding of pattern dynamics. If so it seems unlikely that any comprehensive relationship between developmental morphology and molecular biology of the stem can be developed without both mathematics and biology.

\section*{Statements and Declarations} 
I did not receive support from any organization for this work and have no relevant financial interests to disclose. 
\printbibliography
\section{Document history}
The first version of this document  uploaded to \url{https://arxiv.org/abs/2407.05857} was on 8 July 2024.  Mathematica code used to generate the Figures, along with a version-controlled copy of the source text, can be found at \url{https://github.com/js229/GeometricalPhyllotaxis/tree/radiusfunction}.
\end{document}